**Enhancing thermopower and Nernst signal of high-mobility Dirac carriers by Fermi level tuning in the layered magnet EuMnBi$_2$**

*K. Tsuruda, K. Nakagawa, M. Ochi, K. Kuroki, M. Tokunaga, H. Murakawa, N. Hanasaki, and H. Sakai\**

K. Tsuruda, K. Nakagawa, Prof. M. Ochi, Prof. K. Kuroki, Prof. H. Murakawa, Prof. N. Hanasaki, Prof. H. Sakai
Department of Physics, Osaka University, Toyonaka, Osaka 560-0043, Japan
E-mail: sakai@phys.sci.osaka-u.ac.jp
Prof. M. Tokunaga
The Institute for Solid State Physics, University of Tokyo, Kashiwa, Chiba 277-8581, Japan
Prof. H. Sakai
PRESTO, Japan Science and Technology Agency, Kawaguchi, Saitama 332-0012, Japan

Dirac/Weyl semimetals hosting linearly-dispersing bands have received recent attention for potential thermoelectric applications, since their ultrahigh-mobility carriers could generate large thermoelectric and Nernst power factors. To optimize these efficiencies, the Fermi energy needs to be chemically controlled in a wide range, which is generally difficult in bulk materials because of disorder effects from the substituted ions. Here it is shown that the Fermi energy is tunable across the Dirac point for layered magnet EuMnBi$_2$ by partially substituting Gd$^{3+}$ for Eu$^{2+}$ in the insulating block layer, which dopes electrons into the Dirac fermion layer without degrading the mobility. Clear quantum oscillation observed even in the doped samples allows us to quantitatively estimate the Fermi energy shift and optimize the power factor (exceeding 100 μW/K$^2$cm at low temperatures) in combination with the first-principles calculation. Furthermore, it is shown that Nernst signal steeply increases with decreasing carrier density beyond a simple theoretical prediction, which likely originates from the field-induced gap reduction of the Dirac band due to the exchange interaction with the Eu moments. Thus, the magnetic block layer provides high controllability for the Dirac fermions in EuMnBi$_2$, which would make this series of materials an appealing platform for novel transport phenomena.

# 1. Introduction

In the last 15 years, conducting materials, whose low energy excitation exhibits linear energy-momentum dispersion of relativistic Dirac fermions, have attracted much attention due to their extremely high mobility and unconventional quantum transport properties[1]. While early researches mainly focused on 2D systems, such as graphene[2] and surface states of topological insulators[3], more recently the 3D bulk analogues called Dirac/Weyl semimetals have been intensively explored to discover various exotic physical properties[4, 5]. These include chiral anomaly[6–9], giant magnetoresistance[10, 11], and large anomalous Hall effect[12–16], drawing attention for potential applications to advanced spin-electronic devices. Taking advantage of the bulk form, the Dirac/Weyl semimetals should also be promising for thermoelectric application[17], since the high mobility carriers are able to achieve an excellent power factor as is the case for most of the conventional thermoelectric materials. For Dirac semimetal $Cd_3As_2$, for instance, a dimensionless figure of merit ($ZT$) reaches ~0.2 around 350 K, which is further enhanced to $ZT \sim 1$ by applying magnetic fields up to 7 T owing to the strong magneto-thermopower effect[18]. Furthermore, a recent theoretical study suggested that the low-dimensional (e.g., 1D or anisotropic 2D) gapped Dirac dispersion is even more ideal, because a large density of states and a high group velocity can coexist[19]. In experiments, one-dimensional Dirac semimetal $(Ta,Nb)_4SiTe_4$ was found to show a large power factor at low temperatures[20, 21].

Besides the thermoelectricity, the Dirac linear dispersion could also show high-efficient transverse thermoelectricity, i.e., a large Nernst effect[22, 23], which was so far intensively investigated mainly for Bi[24–26] and graphite[27, 28]. In these materials, the high carrier mobility as well as low carrier density plays a vital role, since the conventional Nernst coefficient roughly tracks $\omega_c\tau/E_F$ [23, 26], where $\omega_c$ is the cyclotron frequency, $\tau$ the relaxation time, and $E_F$ the Fermi energy. In many of the Dirac/Weyl semimetals, such a condition is well satisfied and hence the large Nernst effects were recently observed (e.g., in NbP[29], $Pb_{1-x}Sn_xSe$[30], $Cd_3As_2$ [31, 32], Ta(P,As)[33] and (Zr,Hf)Te$_5$ [34, 35]). In addition to the aforementioned conventional term, the anomalous term arising from the nonzero Berry curvature associated with the Dirac/Weyl points could also contribute to the Nernst signal ($S_{yx}$), when $E_F$ is located close to the Dirac/Weyl point[36, 37]. The rapid increase in $S_{yx}$ at low fields was observed in some materials[31–35], which is likely relevant to the anomalous term.

For the optimization of the thermoelectric and Nernst power factors, wide-range as well as precise tuning of $E_F$ by chemical substitution is indispensable. In 3D bulk materials, however, the substituted ions tend to degrade the carrier mobility and/or modify the peculiar band structure, resulting in a significant reduction in power factor. To overcome this issue, the layer structure consisting of the insulating and conducting layers that are spatially separated is of advantage, since the insulating layer works as a charge reservoir without disturbing the conducting layer (known as the block-layer concept or nano-block integration)[38]. We here propose layered material $A$Mn$X_2$ ($A$: alkaline and rare earth ions, $X$: Sb, Bi)[39–51] can realize this concept in the Dirac/Weyl semimetals, which consists of the alternative stack of the $X^-$ square net layer hosting quasi 2D Dirac fermions[52] and the (Mott) insulating $A^{2+}$-$Mn^{2+}$-$X^{3-}$ block layer (**Figure 1a**). Among them, we focus on EuMnBi$_2$ as a parent material, since the quasi-2D Dirac bands, hosting an energy gap (∼50 meV) due to the strong spin-orbit coupling of Bi[52, 53], cross $E_F$ without contamination of other trivial bands (Figure 1c). This is confirmed by the previous experiments, such as the transport measurement[46] and the photoemission spectroscopy[48]. In this study, we have demonstrated the chemical tuning of $E_F$ across the Dirac point for EuMnBi$_2$ by partially substituting $Eu^{2+}$ with $Gd^{3+}$ to dope electrons with keeping the same $4f^7$ configuration. Reflecting the validity of the block layer concept, the high mobility is retained even for the doped samples, allowing us to quantitatively estimate the $E_F$ shift by observing the quantum oscillation. Based on this, we have experimentally revealed the overall $E_F$ dependence of thermopower ($S_{xx}$) and Nernst signal ($S_{yx}$), which is compared with the theoretical calculation.

## 2. Results and Discussion
### 2.1. Determination of Fermi Energy Shift

We synthesized Gd-doped EuMnBi$_2$ single crystals by changing the nominal Gd concentration of the starting materials (see Methods). Although the energy dispersive x-ray analysis indicated the presence of Gd in the crystals, it was impossible to quantitatively determine the Gd concentration for the low-doped samples by this technique (see Supporting Information). Therefore, we have specified the variation in $E_F$ among the samples based on their transport properties, such as the Shubnikov-de-Haas (SdH) oscillations and Hall effects, as detailed below. We thereby label the Gd-doped samples as Gd#1−#6 in order of increasing $E_F$ (from low to high).

Figures 1d−i show the field dependence of in-plane Hall resistivity $\rho_{yx}$ at 2 K for (Eu,Gd)MnBi$_2$ single crystals. For undoped EuMnBi$_2$, $\rho_{yx}$ has a positive and slightly non-linear slope with respect to field (Figure 1d), indicating the hole-type carriers are doped probably owing to the chemical defects and vacancies. Reflecting the decrease in hole-type carriers (i.e., the increase in $E_F$) with Gd substitution, the slope of $\rho_{yx}$ progressively increases for Gd#1−#3, while it suddenly decreases for Gd#4. It is likely that $E_F$ for Gd#4 is so close to the valence band top that the negative contribution from the electron-type carriers may compensate the positive $\rho_{yx}$ slope. For Gd#5, which is the sample from the same batch as Gd#4, the slope of $\rho_{yx}$ changes to negative, indicating the sign of Hall coefficient is sensitive to a small change in Gd concentration for these samples. For Gd#6, which was grown with the highest nominal Gd concentration, the magnitude of negative $\rho_{yx}$ slope is reduced to ~ 1/100 of that for Gd#5 (inset to Figure 1i), consistent with the further increase in electron-type

carriers. Thus, the measurements of $\rho_{yx}$ have qualitatively revealed the variation of the density and polarity of carriers upon Gd substitution.

It is important that the high mobility for the Dirac fermion on the Bi$^-$ layer remains essentially intact against the Gd substitution in the block layer, leading to the clear SdH oscillations for all the hole-doped samples (Figure 1d-i and insets therein). The increase in $E_F$ with Gd substitution is quantitatively estimated from the variation of the SdH frequency $B_F$, which is proportional to the extremal Femi surface area. For the undoped and Gd#1−#3 samples, the SdH oscillations are extracted from $-\mathrm{d}^2\rho_{yx}/\mathrm{d}(1/B)^2$ at low fields (2.5−5 T) below the spin-flop transition of the Eu layer[46, 54] (insets to Figure 1d−g), whereas that for Gd#4 is discernible only at high fields (6.5−23 T) (inset to Figure 1h. See also Supporting Information). The resultant Landau fan plot is shown in Figure 1b, where the $B_F$ value deduced from the slope systematically decreases from $B_F$ =21 T (undoped) to $B_F$ =9.2 T (Gd#4). In inset to Figure 1b, the calculated $E_F$ versus $B_F$ is denoted by a solid curve, which is roughly expressed by $E_F \propto \sqrt{B_F}$ reflecting the linear dispersion. By plotting the experimental $B_F$ values on this curve, $E_F$ is estimated to shift from -39 meV (undoped) to -25 meV (Gd#4), where $E_F$ is measured from the valence-band top (inset to Figure 1c).

## 2.2. Thermoelectric Performance

Having demonstrated the systematic tuning of $E_F$, we show the corresponding variation of thermoelectric properties for (Eu,Gd)MnBi$_2$ single crystals. **Figure 2** presents the $E_F$ dependence of the temperature profile of in-plane resistivity $\rho_{xx}$ (top panels), $S_{xx}$ (middle), and $S_{yx}$ (bottom) for (Eu,Gd)MnBi$_2$, where $E_F$ increases from Figure 2a (undoped) to Figure 2g (Gd#6). For undoped EuMnBi$_2$, in spite of the nice metallic behavior ($\rho_{xx} \sim$ 90 μΩcm at 300 K), the $S_{xx}$ value is fairly large, reaching the maximum of 70 μV/K at 100 K (middle panel in Figure 2a). Note here that the temperature profile of $S_{xx}$ for normal metals monotonically increases with increasing temperature, which is not the case for (Eu,Gd)MnBi$_2$. The reduction in $S_{xx}$ at high temperatures likely arises from the negative contribution from the thermally-excited carriers in the conduction band, as is reproduced by the first-principles calculations (*vide infra*). When $E_F$ increases for the hole-doped samples, $S_{xx}$ shows an initial increase (Gd#1) followed by a gradual decrease (Gd#2−#4), while leaving the temperature dependence almost unchanged. For Gd#5, $S_{xx}$ changes to negative, while the temperature dependence of the absolute value still remains similar to that for the hole-doped samples. For heavily electron-doped Gd#6, however, $|S_{xx}|$ is much reduced (∼10 μV/K at 300 K), showing a monotonic increase with temperature as is the case for normal metals.

To discuss the detail of the $E_F$ dependence of $S_{xx}$, we shall compare the experimental results with the calculated ones. We here focus on the hole-doped samples, where the carrier density $n$ as well as $E_F$ is quantitatively estimated from the SdH oscillation. Note that $n$ is defined as the summation of the electron density (positive sign) and hole density (negative sign), which are calculated based on the $E_F$ value (Figure 1c). **Figure 3a** shows the $n$ dependence of $S_{xx}$ in the hole-doped region at selected temperatures. At 300 K and 150 K, the experimental data (open squares) are semi-quantitatively reproducible by the first-principles calculation (solid curves), indicating the optimum $n$ value is theoretically predictable. At 75 K, on the other hand, the calculation significantly overestimates the experimental results, especially when $n$ decreases down to ∼ $2.0 \times 10^{19}$ cm$^{-3}$. The decrease in positive $S_{xx}$ value with $E_F$ approaching the Dirac point arises from the negative contribution from the thermally-excited electron-type carriers in the conduction bands, which include not only the Dirac cone but also the parabolic band at the M point (Figure 1c). However, the contribution from the latter does not strongly affect the $n$ dependence of total $S_{xx}$, as was revealed by the additional calculation taking account of the shift of the parabolic band relative to the Dirac point (see Supporting Information). Therefore, the disagreement in $S_{xx}$ can be attributed to the contribution from the conduction band of the Dirac cone, the position of which may not be precisely reproduced by the present calculation.

In addition to this, the $k$-dependence of $\tau$, which is neglected in the constant-$\tau$ approximation, may not be negligible when $E_F$ approaches the Dirac point. To reveal the impact of $\tau$ on $S_{xx}$, the first-principles calculation of $\tau$ is necessary. However, it remains challenging owing to the huge computational cost [55], since the present material requires the spin-orbit coupling and the magnetism to be included in the calculation, otherwise the band gap is closed and the transport calculation fails.

Taking advantage of chemically tunable $S_{xx}$ presented above, we are able to optimize the thermoelectric power factor (PF= $S_{xx}^2/\rho_{xx}$). Figure 3b shows the temperature dependence of PF for the hole-doped samples, where PF exhibits a peak around 50-75 K reflecting the temperature dependence of $S_{xx}$. $\rho_{xx}$ persistently shows nice metallic behavior regardless of the Gd concentration (top panels in Figure 2). For Gd#1-#3, in particular, the residual resistivity ratio [RRR=$\rho_{xx}$(300 K)/$\rho_{xx}$(2 K)], which is a measure of the mean free path ($\propto \tau$), is almost unchanged from the undoped value (top panels in Figure 2a-d and Table S1), demonstrating that the block layer works as an ideal charge reservoir for the Dirac fermion layer. Consequently, the $n$ dependence of PF is similar to that of $S_{xx}$, resulting in the highest PF for Gd#1 at all temperatures. The peak value of PF reaches 105 μW/K$^2$cm at 60 K, which is comparable to that for the best $p$-type thermoelectric materials, such as Na$_x$CoO$_2$ (~160 μW/K$^2$cm at 75 K)[57] and Ti-doped Nb$_4$SiTe$_4$ hosting 1D Dirac-like band (~60 μW/K$^2$cm at 210 K)[21].

### 2.3. Nernst Signal Enhancement toward the Dirac Point

In addition to $S_{xx}$, the Nernst effect is also tunable with Gd substitution, although the $E_F$ dependence is totally different from that of $S_{xx}$ (bottom panels in Figure 2). For the hole-doped samples (Figure 2a-e), the temperature profile of $S_{yx}$ at 9 T exhibits the maximum at some temperature below 100 K, besides a weak anomaly at the antiferromagnetic transition temperature of the Eu layer ($T_N \sim 22$ K). (See also the inset to **Figure 4** for the magnification of the low-temperature profiles) Noteworthy is that the peak value of $S_{yx}$ ($S_{yx}^{max}$) progressively increases as $E_F$ approaches the Dirac point; $S_{yx}^{max}$ reaches 50 µV/K for Gd#3, which is approximately five times as large as $S_{yx}^{max}$ for the undoped sample. For Gd#4 located closest to the Dirac point, however, $S_{yx}^{max}$ suddenly decreases to 24 µV/K, followed by the negligibly small $S_{yx}$ values (< 1.5 µV/K) in the entire temperature range for the electron-doped samples (Figure 2f,g). For a simple 2D Dirac band, the calculation predicts that $S_{yx}$ is an even function of $E_F$, showing a large positive peak at the Dirac point[58, 59], which cannot reproduce the observed electron-hole asymmetric $S_{yx}$. Considering that there are some less-dispersive parabolic conduction bands just above $E_F = 0$ (e.g., at the M point shown in Figure 1c)[48], the vanishingly small $S_{yx}$ for $E_F > 0$ may be caused by the multi-band effect. To explain the $E_F$ dependence of $S_{yx}$ for electron doped samples, therefore, the calculation taking account of the multi-band nature would be necessary, which remains as a future work.

We now discuss the $E_F$ dependence of $S_{yx}$ for the hole-doped samples in details, based on a theoretical model considering a massless Dirac band[29, 60]. It is known that this model reproduces the temperature and field dependence of $S_{yx}$ for Weyl semimetal NbP[29]; specifically, it reproduces the low-temperature peak in $S_{yx}$, as was also observed for the present materials (inset to Figure 4). From the calculation, the peak value and the peak temperature ($T_{max}$) are given by $S_{yx}^{max} \propto v_F^2 \tau / E_F$ and by $T_{max} \propto E_F$, respectively, where $v_F$ is the Fermi velocity. In Figure 4, we plot the experimental $S_{yx}^{max}$ and $T_{max}$ values versus $E_F$, together with the theoretical curves of the above relations. Note here the proportional constants of the theoretical curves are determined so as to reproduce the data for the undoped sample (with the largest $|E_F|$). Although the experimental $T_{max}$ data deviate downward from the theoretical line, its overall $E_F$ dependence roughly tracks $T_{max} \propto E_F$ (Figure 4b). On the other hand, the $S_{yx}^{max}$ data largely deviate upward from the curve of $S_{yx}^{max} \propto 1/E_F$ (Figure 4a). For the linear Dirac band, since $v_F$ is independent of $E_F$, this suggests $\tau$ progressively increases with $E_F$ approaching the Dirac point for Gd#1-#3, followed by a sudden drop for Gd#4. In fact, a marked decrease in $\tau$ for Gd#4 is also suggested in the $\rho_{xx}$ data; the RRR value (a measure of $v_F\tau$) as well as the low-temperature mobility is much reduced for Gd#4 in comparison to the undoped and Gd#1-#3 samples (top panel in Figure 2e and Table S1). Therefore, the decrease in $S_{yx}$ for Gd#4 is likely to arise mainly from the decrease in $\tau$. On the other hand, the RRR value and low-temperature mobility are almost unchanged from the undoped sample to Gd#3 (Table S1). The anomalous enhancement of $S_{yx}^{max}$ for these samples thus cannot be well explained by the simple model adopted here.

**2.4. Impacts of Eu magnetic moments**

As a plausible origin for this, we discuss the impact of the Eu local moments on $S_{yx}$. Above $T_N$ (~22 K), the Eu moments exhibit isotropic paramagnetic behavior following the Curie-Weiss law[46]. Near $T_{max}$ (40-80 K), appreciable net magnetization is induced by the field of 9 T (e.g., 1.7 $\mu_B$/Eu at 50 K), which generates large spin splitting in the gapped Dirac bands owing to the exchange interaction between the Dirac fermions and Eu local moments. From the first-principles calculation[53], the magnitude of spin splitting is roughly estimated to be ~20 meV for 1.7 $\mu_B$/Eu, significantly reducing the energy gap (~50 meV). For a 2D Dirac band, it is theoretically revealed that the non-zero energy gap suppresses the steep increase in $S_{yx}$ toward the Dirac point[58]. Therefore, the magnetization-induced gap reduction may weaken such suppression, leading to the additional enhancement in $S_{yx}^{max}$ observed for Gd#1-#3.

We have also observed a marked impact of the Eu local moments on $S_{xx}$ at 9 T (open circles in the middle panels of Figure 2a-e). The positive $S_{xx}$ values for the hole-doped samples are significantly reduced by the field of 9 T at mid-to-high temperatures. The magnitude as well as temperature range of such negative magneto-thermopower evolves with decreasing $|E_F|$ (from undoped to Gd#4). These facts are again consistent with the band gap reduction at 9 T, since it promotes the negative thermopower from the conduction band. Note here that the negative magneto-thermopower changes to the positive one at low temperatures. The latter can be explained by the Mott's relations within the conventional semiclassical Drude-Boltzmann theory[30], as is often observed in other Dirac/Weyl semimetals with high carrier mobility[30, 31, 34]. For (Eu,Gd)MnBi$_2$, the mobility is enhanced at low temperatures (especially below $T_N$), where the magnitude of the positive magneto-thermopower could overcome the negative one. Thus, the magnetic block layer in the present materials plays a role not only as a source of charge carriers to tune $E_F$, but also as a source of exchange interaction to tune the band gap, which would enable us to improve the thermoelectric and Nernst efficiencies in various ways.

## 3. Conclusion

In conclusion, we have demonstrated the systematic control of Fermi energy for layered Dirac material EuMnBi$_2$ by partially substituting $Gd^{3+}$ for $Eu^{2+}$ in the block layer. Since the high carrier mobility is retained even for the doped samples, we observed the clear quantum oscillation and thereby quantitatively estimate the carrier density and the Fermi energy. Wide-range chemical control of the Fermi energy across the Dirac point enables us to experimentally clarify the overall variation of the transport and thermoelectric properties. The Fermi energy dependence of thermopower is roughly reproduced by the first-principles calculation, achieving the optimized power factor of more than 100 µW/K$^2$cm around 50 K. On the other hand, the Fermi energy dependence of Nernst signal shows an anomalous increase toward the Dirac point, which cannot be explained within a simple rigid-band scheme. The calculation suggests that the field-induced Eu magnetization reduces the gap of the Dirac band via the exchange interaction, which may strongly affect the Nernst signal and thermopower at fields. The present study has demonstrated that a variety of parameters of Dirac fermions are controllable via the magnetic block layer in the $A$Mn$X_2$ materials, offering a platform for novel thermoelectric phenomena associated with the interplay of the topological band and magnetism.

## 4. Experimental Section

*Experimental details*: Single crystals of (Eu,Gd)MnBi$_2$ were grown by a Bi self-flux method. High purity ingots of Eu (99.9%), Gd (99.9%), Mn (99.9%), Bi (99.999%) were mixed in the ratio of Eu : Gd : Mn : Bi = 1-$x$ : $x$ : 1 : 9 and put into an alumina crucible in an argon-filled glove box, where $x$ is 0, 0.005, 0.0075, 0.01, 0.06, and 0.1 for the undoped, Gd#1, Gd#2, Gd#3, Gd#4-#5, and Gd#6 samples, respectively. The crucible was sealed in an evacuated quartz tube and heated at 1000∘C for 10 h, followed by slow cooling to 350∘C at the rate of ~ 2∘C/h, where the excess Bi flux was decanted using a centrifuge. The powder x-ray diffraction profile at room temperature indicates that the crystal structure of the obtained single crystals is tetragonal (*I4/mmm*) and the lattice constants are almost unchanged irrespective of the nominal Gd concentration.

$\rho_{xx}$ and $\rho_{yx}$ were measured by a conventional 5-terminal method with electrodes formed by room-temperature curing silver paste. $S_{xx}$ and $S_{yx}$ were simultaneously measured by a steady-state method with a temperature difference ($\Delta T$) of less than 1 K (typically 2-4% of the measurement temperature below 50 K) between the longitudinal voltage contacts. Note here $S_{yx} = (V_y/\Delta T)(L_x/L_y)$, where $V_y$ is the field-induced transverse voltage, $L_x$ the distance between the longitudinal contacts, and $L_y$ the distance between the transverse contacts. The measurements were performed from 2 K to 300 K at 9 T ($B\|c$), using Physical Properties Measurement System (Quantum Design). For Gd#4, the resistivity up to 23 T at 1.4 K was measured using the non-destructive pulsed magnet with a pulse duration of 36 msec at the International Mega-Gauss Science Laboratory at the Institute for Solid State Physics. The voltage signal was measured by a lock-in technique at 100 kHz with the ac excitation of 1-10 mA.

*Computational details*: We performed first-principles band structure calculations using the density functional theory with the Perdew-Burke-Ernzerhof parametrization of the generalized gradient approximation (PBE-GGA)[61] and the projector augmented wave (PAW) method [62] as implemented in the Vienna *ab initio* simulation package[63–66]. We applied the open-core treatment for the Eu-*f* orbitals, to say, the seven occupied 4*f* orbitals (4$f^7$) for Eu$^{2+}$ ion were included into the core of the PAW potential. The spin-orbit coupling was included. The G-type antiferromagnetic order of Mn atoms as observed in experiment was assumed. The plane-wave cutoff energy of 350 eV and a 16 × 16 × 16 ***k***-mesh were used. Lattice parameters and atomic coordinates were taken from experiment[45].

After the band structure calculation, we extracted the Mn-*d* + Bi-*p* Wannier functions[67, 68] from the calculated band structure using the WANNIER90[69] code. For a technical reason, we used the conventional (tetragonal) unit cell with a 16 × 16 × 4 ***k***-mesh for this purpose. Then, we constructed a tight-binding model with the obtained hopping parameters among the Wannier functions. Using this model, we determined the chemical potential so as to reproduce the experimental values of $B_F$. Here, we evaluated $B_F$ on constant-$k_z$ planes and then took an average with respect to $k_z$.

We also evaluated the carrier density and the Seebeck coefficient using this model and the estimated chemical potential. We calculated the Seebeck coefficient on the basis of the Boltzmann transport theory with a constant *τ* approximation. Note that the Seebeck coefficient does not depend on *τ* if *τ* is assumed to be constant. Here we applied an upward shift for the conduction bands by 30 meV, because PBE-GGA is known to underestimate the band gap. This shift value was determined so that the calculated Seebeck coefficients agree well with experimental ones. These calculations using our tight-binding model were performed using very fine ***k***-meshes from 600 × 600 × 40 to 960 × 960 × 40.

For further calculations shown in Supporting Information, we also used other exchange-correlation functionals. For calculating the band structure shown in Fig. S4(a)-(d), we used an 8 × 8 × 8 ***k***-mesh without including the spin-orbit coupling, because of the high computational cost for hybrid-functional calculations. Here we used the B3LYP[70,71] and HSE06[72] functionals. For the PBE+$U$ calculation shown in Figs. S4 and S5, we adopted the simplified rotationally-invariant formulation introduced by Dudarev *et al.*[73] with $U_{eff} = U - J = 5$ eV for the Mn $d$-orbitals so that the calculated band structure exhibits good consistency with those calculated with the hybrid functionals. For the transport calculation using the PBE+$U$ method shown in Fig. S5, we calculated the band structure including the spin-orbit coupling as shown in Fig. S4(f), extracted the Wannier orbitals, constructed a tight-binding model using these Wannier orbitals, applied the same energy shift as PBE (30 meV) to the conduction bands calculated using this model, and calculated the Seebeck coefficient based on the Boltzmann transport theory. All of these procedures for transport calculation were done using the same computational conditions as the PBE calculation.

**Acknowledgements**
The authors thank S. Ishiwata and H. Masuda for helpful discussions. This work was partly supported by the JST PRESTO (Grant No. JPMJPR16R2), the JSPS KAKENHI (Grant Nos. 19H01851, 19K21851, and 19H05173), the Asahi Glass Foundation, and the IWATANI NAOJI Foundation.

**References**

[1] O. Vafek, A. Vishwanath, Annu. Rev. Condens. Matter Phys. 5, 83 (2014).

[2] A. H. Castro Neto, F. Guinea, N. M. R. Peres, K. S. Novoselov, and A. K. Geim, Rev. Mod. Phys. 81, 109 (2009).

[3] X. L. Qi and S-C. Zhang, Rev. Mod. Phys. 83, 1057 (2011).

[4] N. P. Armitage, E. J. Mele, and A. Vishwanath, Rev. Mod. Phys. 90, 015001 (2018).

[5] B. Yan and C. Felser, Annu. Rev. Condens. Matter Phys. 8, 337 (2017).

[6] X. Huang, L. Zhao, Y. Long, P. Wang, D. Chen, Z. Yang, H. Liang, M. Xue, H. Weng, Z. Fang, X. Dai, and G. Chen, Phys. Rev. X 5, 031023 (2015).

[7] J. Xiong, S. K. Kushwaha, T. Liang, J. W. Krizan, M. Hirschberger, W. Wang, R. J. Cava, and N. P. Ong, Science 350, 413-416 (2015).

[8] M. Hirschberger, S. Kushwaha, Z. Wang, Q. Gibson, S. Liang, C. A. Belvin, B. A. Bernevig, R. J. Cava and N. P. Ong, Nat. Mater. 15, 1161 (2016).

[9] K. Kuroda, T. Tomita, M.-T. Suzuki, C. Bareille, A. A. Nugroho, P. Goswami, M. Ochi, M. Ikhlas, M. Nakayama, S. Akebi, R. Noguchi, R. Ishii, N. Inami, K. Ono, H. Kumigashira, A. Varykhalov, T. Muro, T. Koretsune, R. Arita, S. Shin, T. Kondo and S. Nakatsuji, Nat. Mater. 16, 1090 (2017).

[10] T. Liang, Q. Gibson, M. N. Ali, M. Liu, R. J. Cava, and N. P. Ong, Nat. Mater. 14, 280 (2015).

[11] C. Shekhar, A. Nayak, Y. Sun, M. Schmidt, M. Nicklas, I. Leermakers, U. Zeitler, Y. Skourski, J. Wosnitza, Z. Liu, Y. Chen, W. Schnelle, H. Borrmann, Y. Grin, C. Felser, and B. Yan, Nat. Phys. 11, 645 (2015).

[12] S. Nakatsuji, N. Kiyohara, and T. Higo, Nature 527 212 (2015).

[13] A. K. Nayak, J. E Fischer, Y. Sun, B. Yan, J. Karel, A. C. Komarek, C. Shekhar, N. Kumar, W. Schnelle, J. Kubler, C. Felser, and S. S. P. Parkin, Sci. Adv. 2, e1501870 (2016).

[14] T. Suzuki, R. Chisnell, A. Devarakonda, Y.-T. Liu, W. Feng, D. Xiao, J. W. Lynn, and J. G. Checkelsky, Nat. Phys. 12, 1119 (2016).

[15] T. Liang, J. Lin, Q. Gibson, S. Kushwaha, M. Liu, W. Wang, H. Xiong, J. A. Sobota, M. Hashimoto, P. S. Kirchmann, Z-X. Shen, R. J. Cava, and N. P. Ong., Nat. Phys. 14, 451 (2018).

[16] E. Liu, Y. Sun, N. Kumar, L. Muechler, A. Sun, L. Jiao, S-Y. Yang, D. Liu, A. Liang, Q. Xu, J. Kroder, V. S üß, H. Borrmann, C. Shekhar, Z. Wang, C. Xi, W. Wang, W. Schnelle, S. Wirth, Y. Chen, S. T. B. Goennenwein, and C. Felser., Nat. Phys. 14, 1125 (2018).

[17] C. Fu , Y. Sun, and C. Felser, APL Mater. 8, 040913 (2020)

[18] H. Wang, X. Luo, W. Chen, N. Wang, B. Lei, F. Meng, C. Shang, L. Ma, T. Wu, X. Dai, Z. Wang, and X. Chen, Sci. Bull. 63, 411 (2018).

[19] M. Ochi, H. Usui, and K. Kuroki, Phys. Rev. Applied 8, 064020 (2017).

[20] T. Inohara, Y. Okamoto, Y. Yamakawa, A. Yamakage, and K. Takenaka, Appl. Phys. Lett. 110, 183901 (2017).

[21] Y. Okamoto, Y. Yoshikawa, T. Wada, and K. Takenaka, Appl. Phys. Lett. 115, 043901 (2019).

[22] A. v. Ettingshausen and W. Nernst, Ann. Phys. Chem. 265, 343 (1886).

[23] K. Behnia and H. Aubin, Rep. Prog. Phys. 79, 046502 (2016).

[24] W. M. Yim, and A. Amith, Solid-State Electron. 15, 1141 (1972).

[25] K. Sugihara, J. Phys. Soc. Jpn. 27, 362-370 (1969).

[26] Kamran Behnia, Marie-Aude Measson, and Yakov Kopelevich, Phys. Rev. Lett. 98, 076603 (2007).

[27] K. Sugihara, T. Takezawa, T. Tsuzuku, Y. Hishiyama, and A. Ono, J. Phys. Chem. Solids 33, 1475 (1972).

[28] Z. Zhu, H. Yang, B. Fauque, Y. Kopelevich and K. Behnia . Nat. Phys. 6, 26 (2010).

[29] S. J. Watzman, T. M. McCormick, C. Shekhar, S.C. Wu, Y. Sun, A. Prakash, C. Felser, N. Trivedi, and J. P. Heremans, Phys. Rev. B 97, 161404(R) (2018).

[30] T. Liang, Q. Gibson, J. Xiong, M. Hirschberger, S. P. Koduvayur, R. J. Cava, and N. P. Ong, Nat. Commun. 4, 2696 (2013).

[31] T. Liang, J. Lin, Q. Gibson, T. Gao, M. Hirschberger, M. Liu, R. J. Cava, and N. P. Ong, Phys. Rev. Lett. 118, 136601 (2017).

[32] J. Xiang, S. Hu, M. Lyu, W. Zhu, C. Ma, Z. Chen, F. Steglich, G. Chen, and P. Sun, Sci. China: Phys., Mech. Astron. 63, 237011 (2019).

[33] F. Caglieris, C. Wuttke, S. Sykora, V. Suβ, C. Shekhar, C. Felser, B. Buchner, and C. Hess, Phys. Rev. B 98, 201107(R) (2018).

[34] J. L. Zhang, C. M. Wang, C. Y. Guo, X. D. Zhu, Y. Zhang, J. Y. Yang, Y. Q. Wang, Z. Qu, L. Pi, Hai-Zhou Lu, and M. L. Tian, Phys. Rev. Lett. 123, 196602 (2019).

[35] J. Hu, M. Caputo, E. Bonini Guedes, Sa Tu, E. Martino, A. Magrez, H. Berger, J. Hugo Dil, H. Yu, and J.-P. Ansermet, Phys. Rev. B 100, 115201 (2019).

[36] D. Xiao, Y. Yao, Z. Fang, and Q. Niu, Phys. Rev. Lett. 97, 026603 (2006).

[37] G. Sharma, P. Goswami, and S. Tewari, Phys. Rev. B 93, 035116 (2016).

[38] K. Koumoto, I. Terasaki, and R. Funahashi, MRS Bull. 31, 206 (2006).

[39] J. Park, G. Lee, F. Wolff-Fabris, Y. Y. Koh, M. J. Eom, Y. K. Kim, M. A. Farhan, Y. J. Jo, C. Kim, J. H. Shim, and J. S. Kim, Phys. Rev. Lett. 107, 126402 (2011).

[40] J. K. Wang, L. L. Zhao, Q. Yin, G. Kotliar, M. S. Kim, M. C. Aronson, and E. Morosan, Phys. Rev. B 84, 064428 (2011).

[41] K. Wang, D. Graf, H. Lei, S. W. Tozer, and C. Petrovic, Phys. Rev. B 84, 220401(R) (2011).

[42] L.-L. Jia, Z.-H. Liu, Y.-P. Cai, T. Qian, X.-P. Wang, H. Miao, P. Richard, Y.-G. Zhao, Y. Li, D.-M. Wang, J.-B. He, M. Shi, G.-F. Chen, H. Ding, and S.-C. Wang, Phys. Rev. B 90, 035133 (2014).

[43] L. Li, K. Wang, D. Graf, L. Wang, A. Wang, and C. Petrovic, Phys. Rev. B 93, 115141 (2016).

[44] Y. Y. Wang, Q. H. Yu and T. L. Xia, Chinese Phys. B 25, 107503 (2016).

[45] A. F. May, M. A. McGuire, and B. C. Sales, Phys. Rev. B 90, 075109 (2014).

[46] H. Masuda, H. Sakai, M. Tokunaga, Y. Yamasaki, A. Miyake, J. Shiogai, S. Nakamura, S. Awaji, A. Tsukazaki, H. Nakao, Y. Murakami, T. Arima, Y. Tokura, and S. Ishiwata, Sci. Adv. 2, e1501117 (2016).

[47] A. Wang, I. Zaliznyak, W. Ren, L. Wu, D. Graf, V. O. Garlea, J. B. Warren, E. Bozin, Yimei Zhu, and C. Petrovic, Phys. Rev. B 94, 165161 (2016).

[48] S. Borisenko, D. Evtushinsky, Q. Gibson, A. Yaresko, K. Koepernik, T. Kim, M. N. Ali, J. van den Brink, M. Hoesch, A. Fedorov, E. Haubold, Y. Kushnirenko, I. Soldatov, R. Schafer and R. J. Cava, Nat. Commun. 10, 3424 (2019).

[49] J. Liu, J. Hu, H. Cao, Y. Zhu, A. Chuang, D. Graf, D. J. Adams, S. M. A. Radmanesh, L. Spinu, I. Chiorescu, and Z. Mao, Sci. Rep. 6, 30525 (2015).

[50] S. Huang, J. Kim, W. A. Shelton, E. W. Plummer, and R. Jin, Proc. Natl. Acad. Sci. USA, 114, 6256 (2017).

[51] H. Sakai, H. Fujimura, S. Sakuragi, M. Ochi, R. Kurihara, A. Miyake, M. Tokunaga, T. Kojima, D. Hashizume, T. Muro, K. Kuroda, Takeshi Kondo, T. Kida, M. Hagiwara, K. Kuroki, M. Kondo, K. Tsuruda, H. Murakawa, and N. Hanasaki, Phys. Rev. B 101, 081104(R) (2020)

[52] G. Lee, M. A. Farhan, J. S. Kim, and J. H. Shim, Phys. Rev. B 87, 245104 (2013).

[53] H. Masuda, H. Sakai, M. Tokunaga, M. Ochi, H. Takahashi, K. Akiba, A. Miyake, K. Kuroki, Y. Tokura, and S. Ishiwata, Phys. Rev. B 98, 161108(R) (2018).

[54] There is a weak anomaly in $\rho_{yx}$ at the field of spin-flop transition of the Eu layer (∼5.3 T), denoted by the filled triangles in Figure 1d-g. To avoid this, we adopted the low-field region of 2.5−5 T.

[55] For a review, F. Giustino, Rev. Mod. Phys. **89**, 015003 (2017).

[56] H. Masuda, H. Sakai, H. Takahashi, Y. Yamasaki, A. Nakao, T. Moyoshi, H. Nakao, Y. Murakami, T. Arima, and S. Ishiwata, Phys. Rev. B 101, 174411 (2020).

[57] M. Lee, L. Viciu, Y. Wang, M. L. Foo, S. Watauchi, R. A. Pascal, Jr., R. J. Cava, and N. P. Ong, Nat. Mater. 5, 537 (2006).

[58] V. P. Gusynin and S. G. Sharapov, Phys. Rev. B 73, 245411 (2016).

[59] A. Kundu, M. A. Alrefae, and T. S. Fisher, J. Appl. Phys. 121, 125113 (2017).

[60] G. Sharma, C. Moore, S. Saha, and S. Tewari, Phys. Rev. B 96, 195119 (2017).

[61] J. P. Perdew, K. Burke, and M. Ernzerhof, Phys. Rev. Lett. 77, 3865 (1996).

[62] G. Kresse and D. Joubert, Phys. Rev. B 59, 1758 (1999).

[63] G. Kresse and J. Hafner, Phys. Rev. B 47, 558(R) (1993).

[64] G. Kresse and J. Hafner, Phys. Rev. B 49, 14251 (1994).

[65] G. Kresse and J. Furthmuller, Comput. Mater. Sci. 6, 15 (1996).

[66] G. Kresse and J. Furthmuller, Phys. Rev. B 54, 11169 (1996).

[67] N. Marzari and D. Vanderbilt, Phys. Rev. B 56, 12847 (1997).

[68] I. Souza, N. Marzari, and D. Vanderbilt, Phys. Rev. B 65, 035109 (2001).

[69] A. A. Mostofi, J. R. Yates, Y.-S. Lee, I. Souza, D. Vanderbilt, and N. Marzari, Comput. Phys. Commun. 178, 685 (2008).

[70] A. D. Becke, J. Chem. Phys. 98 5648 (1993).

[71] P. J. Stephens, F. J. Devlin, C. F. Chabalowski, and M. J. Frisch, J. Phys. Chem. 98 11623 (1994).

[72] A. V. Krukau, O. A. Vydrov, A. F. Izmaylov, and G. E. Scuseria, J. Chem. Phys. 125, 224106 (2006).

[73] S. L. Dudarev, G. A. Botton, S. Y. Savrasov, C. J. Humphreys, and A. P. Sutton, Phys. Rev. B 57, 1505 (1998).

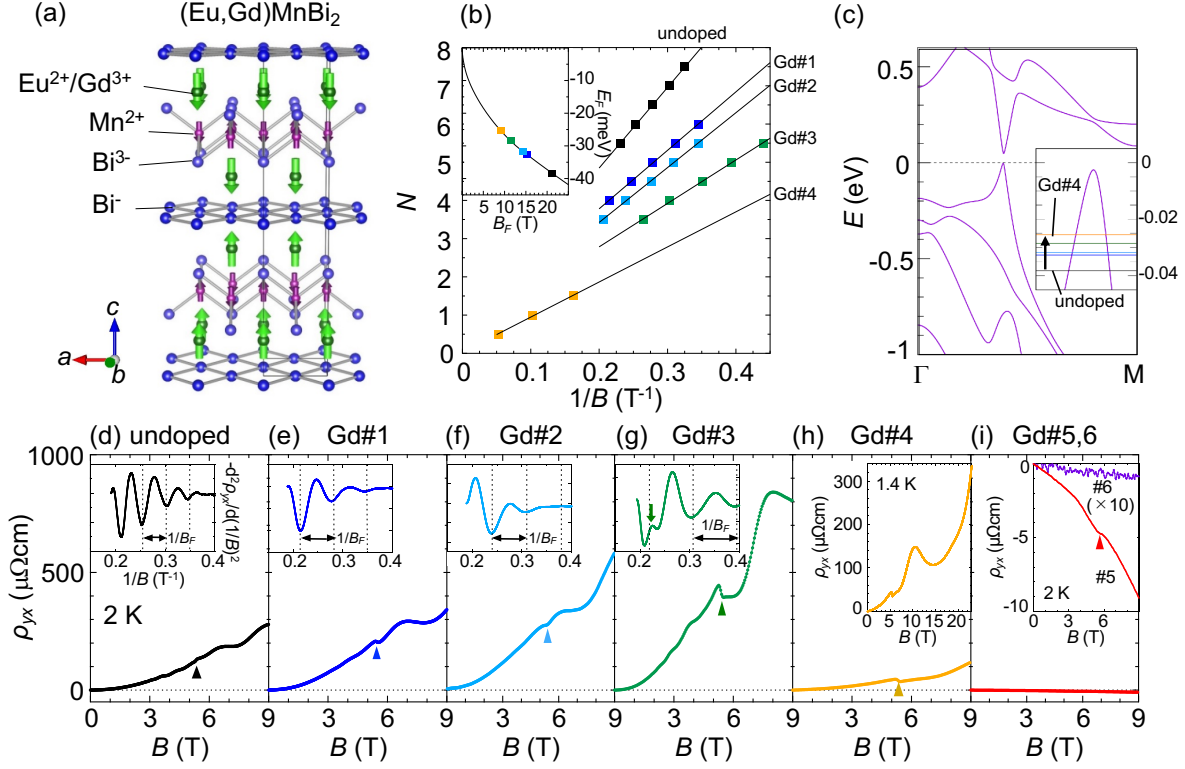

**Figure 1.** Fermi level control by Gd substitution in EuMnBi$_2$. a) Schematic illustration of the crystal and magnetic structure at 0 T for EuMnBi$_2$[56], where Eu$^{2+}$ ions are partially substituted with Gd$^{3+}$ ions. b) Landau fan diagram deduced from the SdH oscillation of the Hall resistivity [insets to panels (d)-(h)]. The inset shows the calculated Fermi energy $E_F$ versus the SdH frequency $B_F$ (solid curve) together with the experimental data (solid squares). c) Calculated band structure along the Γ-M line for EuMnBi$_2$. The inset shows a magnified view of the valence band of the gapped Dirac cone, where the position of $E_F$ for each sample is shown. d)−i) Hall resistivity $\rho_{yx}$ at 2 K versus field $B$ for (Eu,Gd)MnBi$_2$ single crystals ($B\|c$). The filled triangle denotes the field of spin-flop transition of the Eu layer[55]. The inset to (d)−(g) shows $-d^2\rho_{yx}/d(1/B)^2$ versus $1/B$ to estimate $B_F$. The splitting of the oscillation denoted by a vertical arrow in (g) stems from the spin-split Landau levels[53]. The inset to (h) shows $\rho_{yx}$ up to 25 T for Gd#4. The inset to (i) shows a magnified view of $\rho_{yx}$ for Gd#5 and Gd#6.

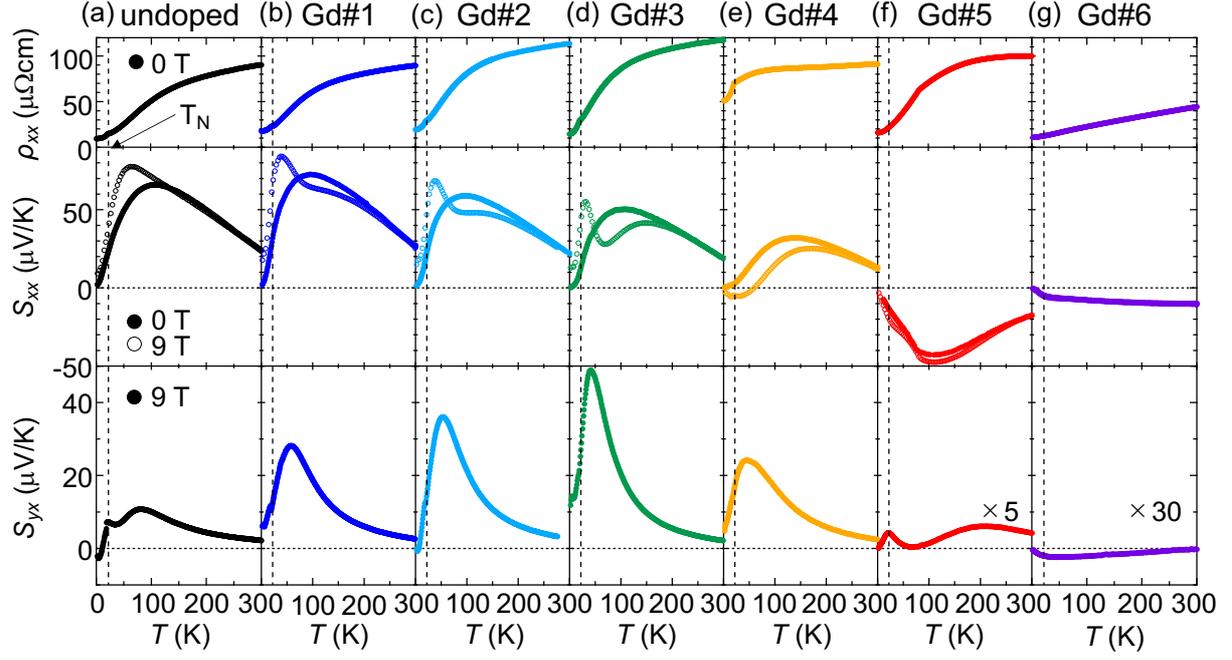

**Figure 2.** Variation of thermoelectric properties upon $E_F$ for (Eu,Gd)MnBi$_2$ single crystals. a)-g) In-plane resistivity $\rho_{xx}$ (top), thermopower $S_{xx}$ (middle), and Nernst signal $S_{yx}$ (bottom) versus temperature T. The magneto-thermopower and Nernst signal were measured at 9 T ($B\|c$). A vertical dashed line denotes the antiferromagnetic transition temperature of the Eu layer ($T_N$) for each sample.

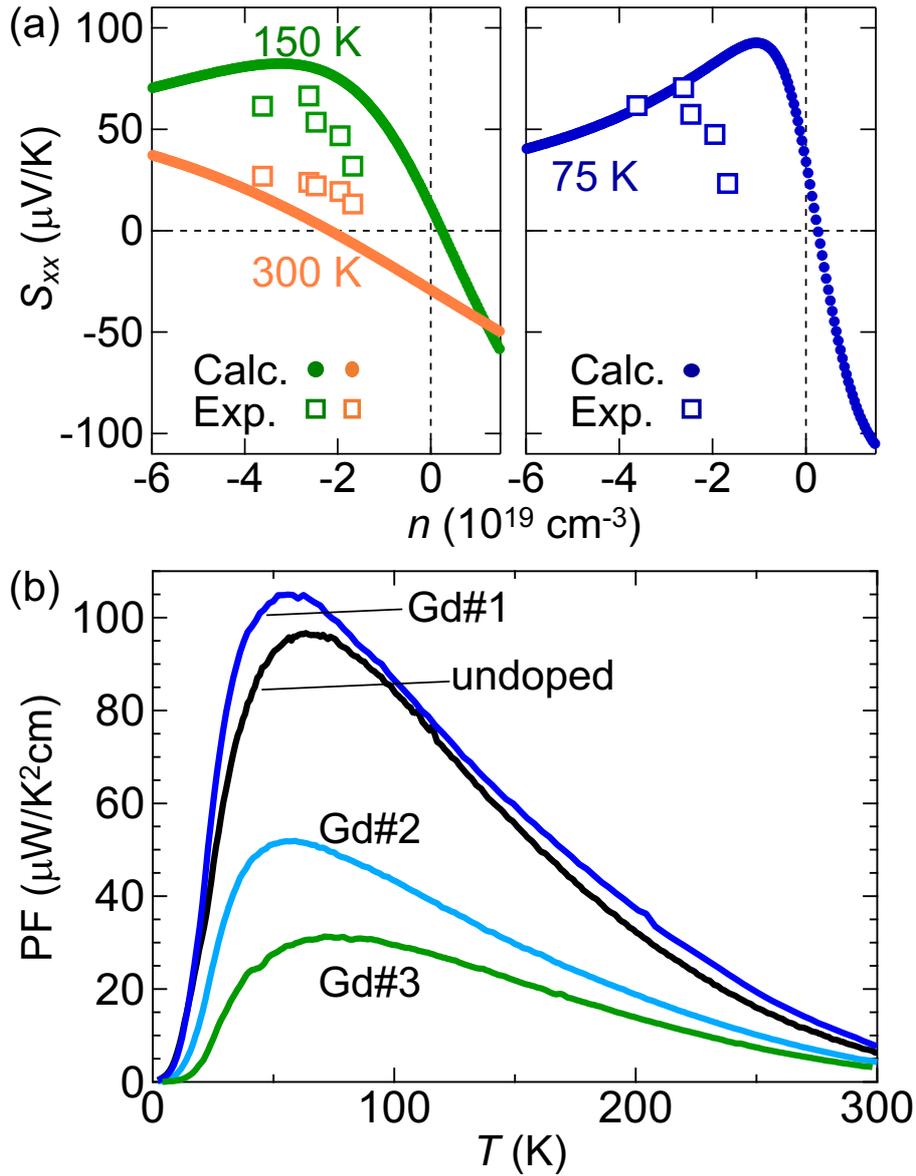

**Figure 3.** Thermoelectric efficiency tunable with the carrier density. a) Thermopower $S_{xx}$ versus carrier density $n$ at selected temperatures. The calculated results are denoted by closed circles, while the experimental data for the hole-doped samples (i.e., undoped to Gd#4) are denoted by open squares. The negative (positive) sign of $n$ indicates the hole-type (electron-type) carrier. The experimental carrier density is estimated from $E_F$ (see Supporting Information). b) Thermoelectric power factor PF versus temperature $T$ for the hole-doped samples.

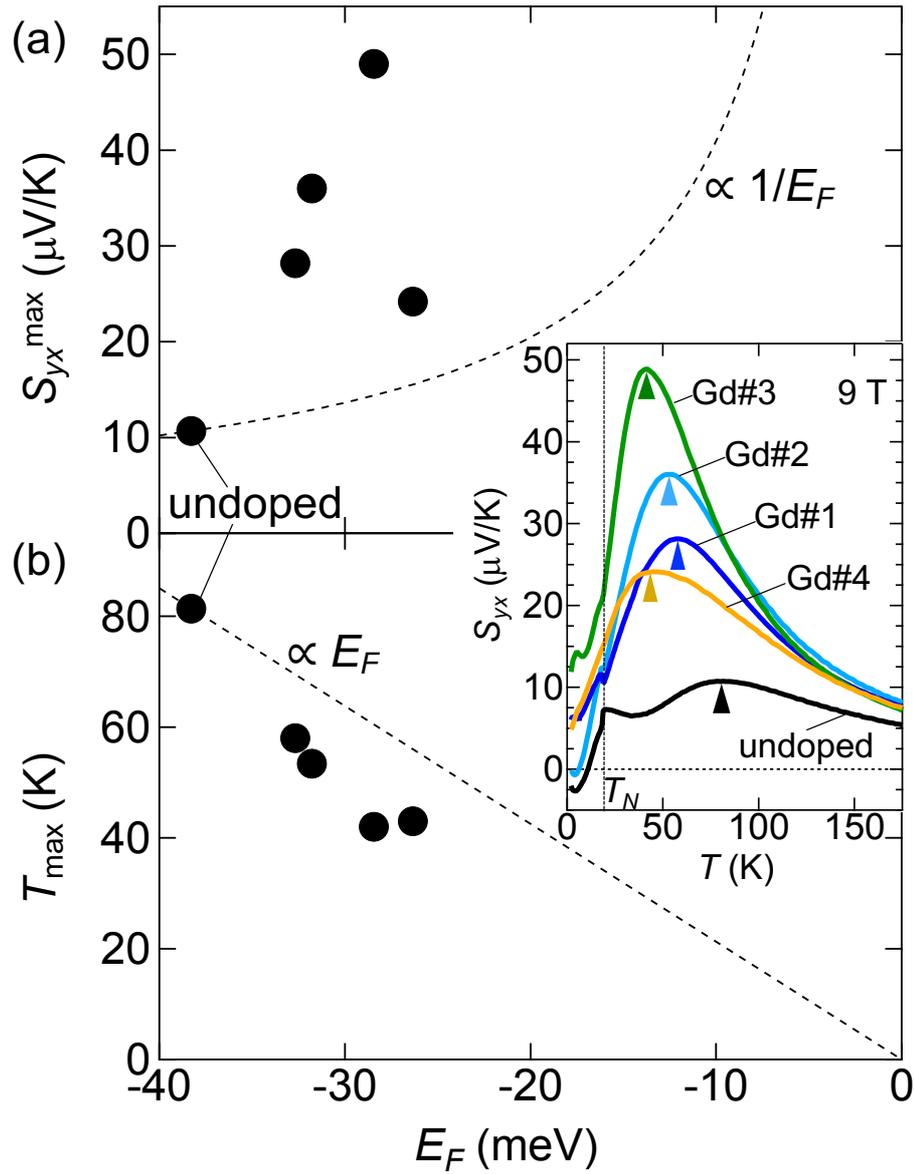

**Figure 4.** Variation of Nernst signal upon Fermi energy. $E_F$ dependence of a) the peak value of $S_{yx}$ at 9 T ($S_{yx}^{max}$) and (b) the peak temperature ($T_{max}$) for hole-doped (Eu,Gd)MnBi$_2$ ($E_F < 0$). The dotted curve in each panel denotes the theoretical one deduced from the data for the undoped sample (see the main text for details). Inset shows the magnification of the low-temperature profile of $S_{yx}$ at 9 T, where the filled triangle corresponds to the peak position for each sample.

# Supporting Information

**Enhancing thermopower and Nernst signal of high-mobility Dirac carriers by Fermi level tuning in the layered magnet EuMnBi$_2$**

*K. Tsuruda, K. Nakagawa, M. Ochi, K. Kuroki, M. Tokunaga, H. Murakawa, N. Hanasaki, and H. Sakai\**

**Energy dispersive x-ray analysis (EDX)**

Figures S1 a and b show the spectra of EDX for Gd#6 (with the highest nominal Gd concentration) and undoped samples, respectively. For Gd#6, when compared to the spectra for the undoped one, weak hump structures are barely discernible at the Gd *L* edges (denoted by orange arrows in Fig. S1a), indicating the presence of Gd. The resultant chemical composition for Gd#6 is give by Eu : Gd : Mn : Bi = 27.52 : 1.99 : 26.66 : 43.83. This suggests that approximately 7% Eu is substituted by Gd, which is smaller than the nominal value (10%). The similar EDX spectrum was observed for the single crystal from the same batch as Gd#4 and Gd#5. Its chemical composition is given by Eu : Gd : Mn : Bi = 25.31 : 0.84 : 26.63 : 47.22, corresponding to approximately 3% Gd substitution. However, because of the peak overlap of Eu and Gd and the small amount of Gd, the error bar of Gd concentration is significant for both measurements (e.g. 7% ± 3% for Gd#6). For the samples with fewer Gd nominal concentration (Gd#1-#3), it is further difficult to quantitatively estimate the actual Gd concentration by the EDX.

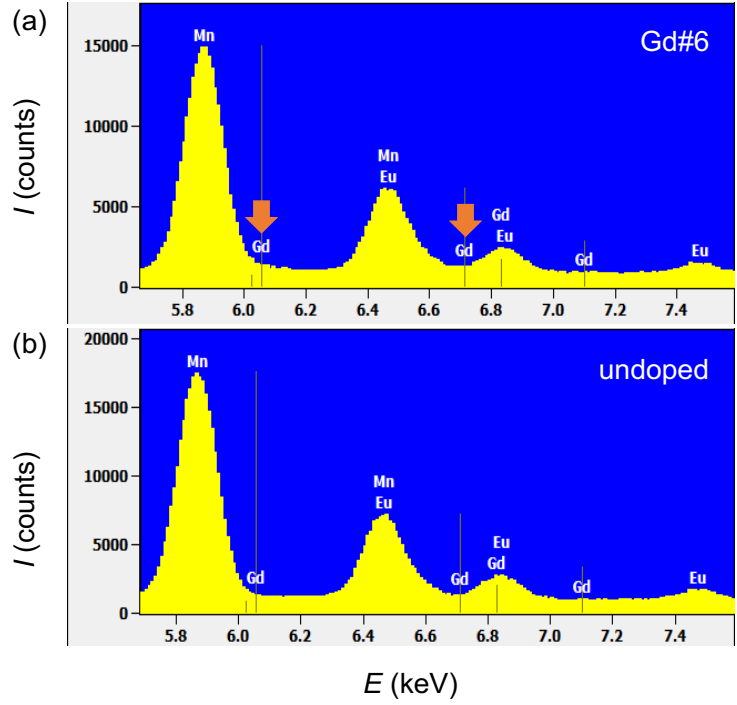

**Figure S1.** Energy dispersive x-ray analysis. The spectra for a) Gd#6 and b) undoped samples are shown. The energy range near the Gd $L$ edge is selected.

**Analysis of the SdH oscillation for Gd#4**

To estimate the SdH frequency $B_F$ for Gd#4, we measured the in-plane resistivity $\rho_{xx}$ and Hall resistivity $\rho_{yx}$ up to 23 T at the lowest temperature (1.4 K) using a pulsed magnet (**Figure S2a,b**), since the SdH oscillation was not observed below the field of spin-flop transition ($H_f$ = 5.3 T). Based on these data, we calculated the in-plane conductivity $\sigma_{xx}$ (Figure S2c), whose SdH oscillation should reflect the density of states of the Landau levels. To extract the oscillation component, we plot $-d^2\sigma_{xx}/d(1/B)^2$ versus $1/B$ (Figure S2), where a deep minimum at $1/B \approx 0.1$ corresponds to the gap between the Landau levels. In addition to this minimum, we see weak dip structures in $-d^2\sigma_{xx}/d(1/B)^2$ (denoted by filled triangles), which are likely to arise from the splitting of the Landau levels. In fact, it was experimentally and theoretically revealed for undoped EuMnBi$_2$ that the field-induced Eu magnetization (above $H_f$) causes significant spin splitting of the Dirac bands via the exchange interaction (see Ref. [53] in the main text). By assuming that the position of the dip structure due to the spin-splitting correspond to the middle of each Landau level, we have constructed the Landau fan diagram (Figure S2e), where the $B_F$ value is estimated to be $B_F = 9.2 \pm 0.9$ T.

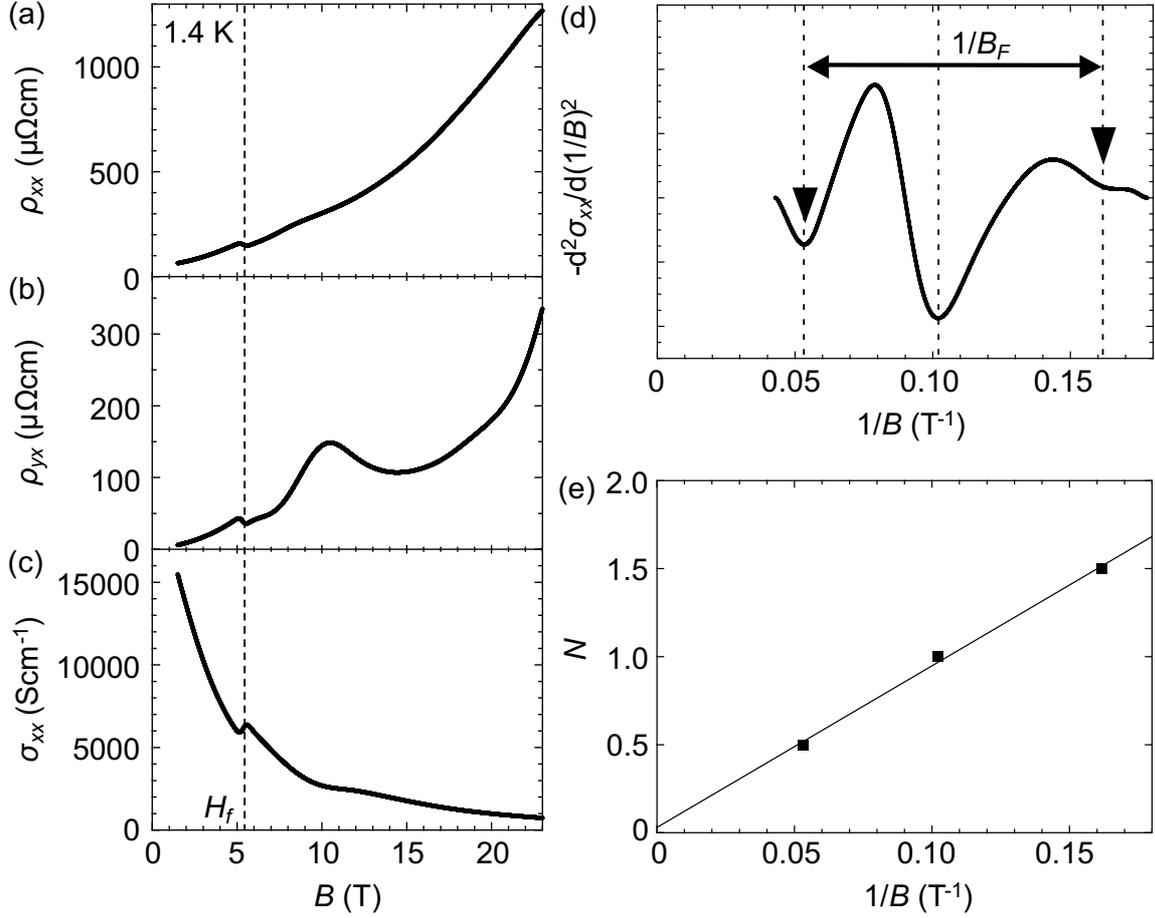

**Figure S2.** SdH oscillation for Gd#4. a) In-plane resistivity $\rho_{xx}$, b) Hall resistivity $\rho_{yx}$, and c) in-plane conductivity $\sigma_{xx}$ versus field $B$ up to 23 T at 1.4 K. $H_f$ (= 5.3 T) denotes the spin-flop transition field of the Eu layer. d) $-d^2\sigma_{xx}/d(1/B)^2$ versus $1/B$ to estimate the SdH frequency ($B_F$). (e) Landau fan diagram deduced from $\sigma_{xx}$.

**Thermal conductivity and dimensionless figure of merit ($ZT$)**

Figures S3 a and b show the temperature dependence of in-plane thermal conductivity $\kappa_{xx}$ and dimensionless figure of merit $ZT =(PF/ \kappa_{xx})T$ for hole-doped EuMnBi$_2$ single crystals (undoped, Gd#1, and Gd#3), respectively. Reflecting the small amount of Gd substitution, the $\kappa_{xx}$ values are similar for the undoped and Gd#3 samples. Since the crystal size for Gd#1 is so small to precisely measure the $\kappa_{xx}$ value, we here adopted the $\kappa_{xx}$ value for Gd#3 to calculate the $ZT$ value for Gd#1. The $ZT$ value shows a peak at 120-140 K, which reaches ~0.09 for Gd#1 (at ~130 K).

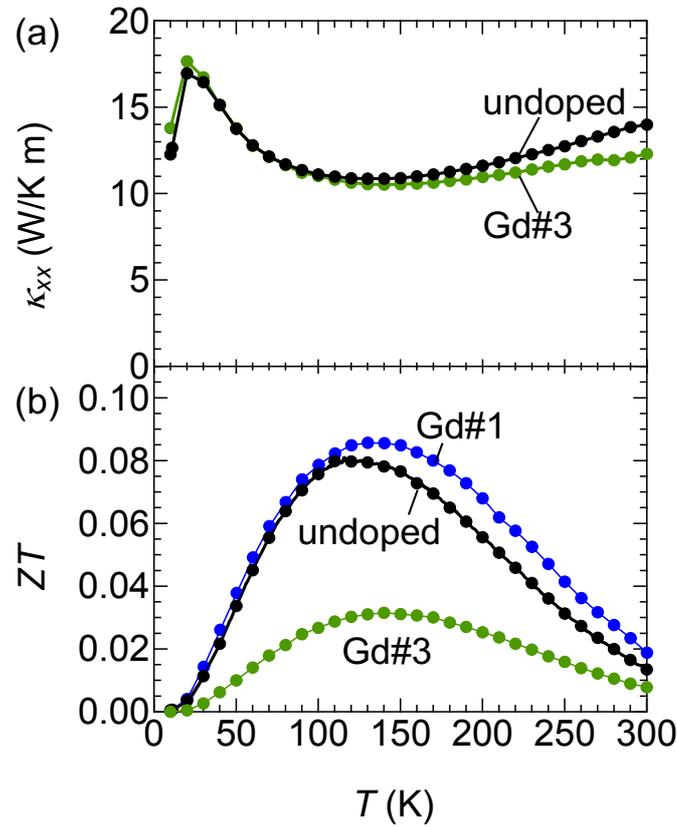

**Figure S3.** Thermoelectric efficiency for hole-doped EuMnBi$_2$. Temperature dependence of a) in-plane thermal conductivity $\kappa_{xx}$ and b) dimensionless figure of merit $ZT$.

**List of the RRR value and mobility at 2 K for hole-doped EuMnBi$_2$**

**Table S1.** The RRR value [$=\rho_{xx}(300\ \text{K})/\rho_{xx}(2\ \text{K})$] and mobility at 2 K for hole-doped samples.

|  | undoped | Gd#1 | Gd#2 | Gd#3 | Gd#4 |
|---|---|---|---|---|---|
| RRR | 9.0 | 5.3 | 6.0 | 8.4 | 1.8 |
| Mobility at 2 K (cm$^2$/Vs) | 18,000 | 14,000 | 13,400 | 22,000 | 7,300 |

**Band structures and thermopower obtained by other calculation methods**

To check the reliability of the calculated band structure shown in the main text, we also calculated the band structure using hybrid functionals, which often have better accuracy than PBE. In Fig. S4(a)-(c), we present the band structures calculated with PBE and popular hybrid functionals: B3LYP and HSE06. Because of the high computational cost for hybrid functionals, we did not include the spin-orbit coupling here. It is found that the conduction band bottom at the M point, which lies relatively close to the Dirac point for PBE, goes higher in energy for the hybrid functionals. Considering better reliability of hybrid functionals, it seems that this conduction band unlikely has a large contribution to thermopower in real materials where the hole carriers are doped. We also found that the band structure calculated with the hybrid functionals is well reproduced by the PBE+$U$ method as shown in Fig. S4(d). We adopted $U_{eff} = U - J = 5$ eV for the Mn $d$-orbitals so that the calculated band structure shows good consistency with those calculated with the hybrid functionals. Because the computational cost of the PBE+$U$ method is much cheaper than that for the hybrid functionals, we used the PBE+$U$ method for further analysis shown here. We also calculated the band structure with including the spin-orbit coupling for PBE and PBE+$U$ methods as shown in Fig. S4(e)-(f). The conduction band at the M point goes higher energy in the PBE+$U$ calculation, even when the spin-orbit coupling is included.

Given these band calculations, we checked that the conduction band at the M point affects little the calculated thermopower [Fig. 3(a) in the main text]. For this purpose, we first calculated the band structure with the PBE+$U$ method including the spin-orbit coupling as shown in Fig. S4(f). Using this band structure, we calculated thermopower in the same way as was performed for PBE (see computational details in experimental section). The calculation results are shown in Fig. S5. Because the calculated thermopower is similar for the PBE and PBE+$U$ methods, we can conclude that thermopower calculated by PBE shown in Fig. 3(a) in the main text is hardly affected by the conduction band at the M point. Therefore, it is suggested that the disagreement between theory and experiment mainly originates from the property of the Dirac band.

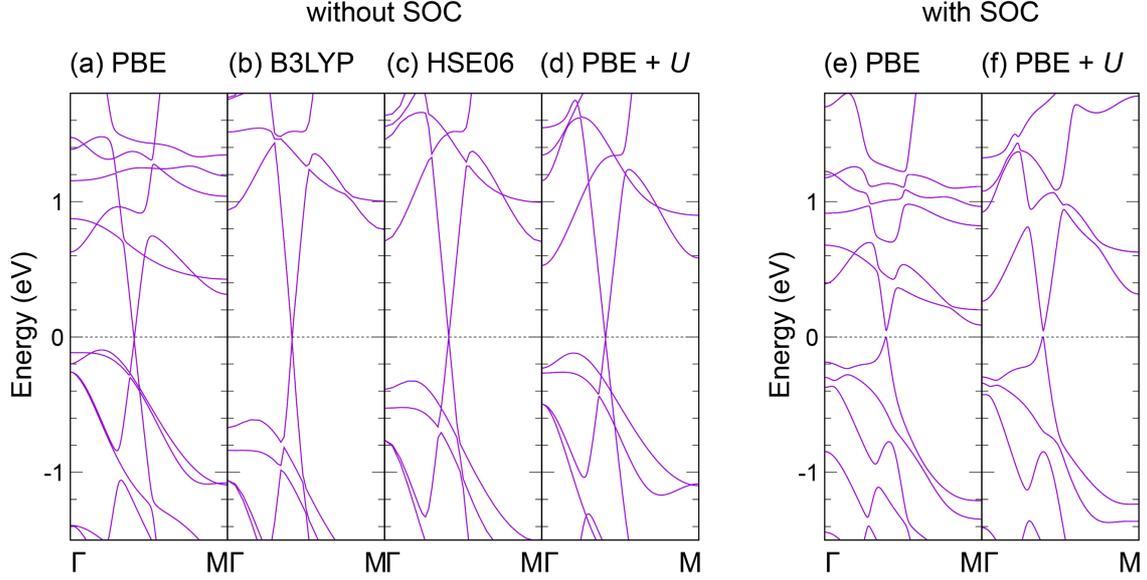

**Figure S4.** The first-principles band structures calculated with (a) PBE, (b) B3LYP, (c) HSE06, and (d) PBE+$U$ methods without including the spin-orbit coupling. The band structures calculated including the spin-orbit coupling are also shown for (e) PBE and (f) PBE+$U$ methods. For comparison among the different methods, we set the energy zero to be the Dirac point along the Γ-M line for panels (a)-(d) and the valence band top along the Γ-M line for panels (e)-(f).

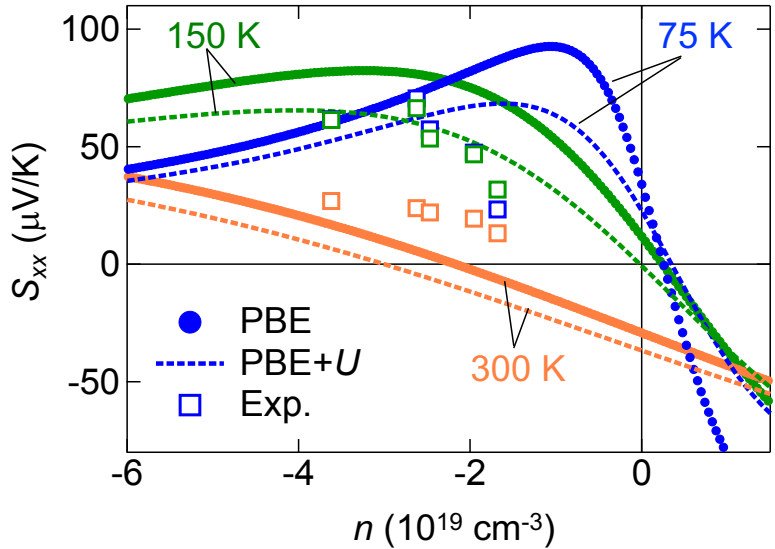

**Figure S5.** Thermopower ($S_{xx}$) versus carrier density ($n$) calculated with the PBE and PBE+$U$ methods together with the experimental values (Exp.) shown in Fig. 3(a) in the main text. Spin-orbit coupling is included in transport calculations shown here.